\documentclass[final]{meson06}
 \usepackage{pifont}   
  \usepackage{bm}

\usepackage{color}
 
  \newcommand{\BluTn}[1]{\textcolor{blue}{#1}}
   \newcommand{\RedTn}[1]{\textcolor{red}{#1}}
\newcommand{\va}[1]{\langle{#1}\rangle} 
\begin{document}
\markboth{\textsc{Bakulev and Pimikov:~Pion quark structure in QCD}}{\textsc{Bakulev and Pimikov:~Pion quark structure in QCD}}
%
\catchline{}{}{}{}{}
%
\title{Pion quark structure in QCD}

\author{A.~P.~Bakulev$^\dag$ and A.~V.~Pimikov$^\ddag$}
\address{Bogolyubov Lab. Theor. Phys., JINR, Dubna, 141980 Russia\\
         $^\dag$bakulev@theor.jinr.ru;~~~$^\ddag$pimikov@theor.jinr.ru}

\maketitle      

\begin{abstract}
 We describe the present status of the pion distribution amplitude
 as it originated from two sources:
 (i) a nonperturbative approach, based on QCD sum rules with nonlocal
 condensates and
 (ii) a NLO QCD analysis of the CLEO data on
 $F^{\gamma\gamma^*\pi}(Q^2)$, supplemented 
 by the recent high-precision lattice calculations 
 of the second moment of the pion distribution amplitude.
\end{abstract}

\ccode{PACS numbers: 12.38.Aw,12.38.Cy,12.38.Lg,13.40.Gp}

\section{Pion distribution amplitude from QCD sum rules}
The pion distribution amplitude (DA), $\varphi_{\pi}$,~\cite{Rad77}
\begin{eqnarray}
 \langle{0\mid \bar d(z)\gamma_{\mu}\gamma_5
 {\cal E}(z,0) u(0)\mid \pi(P)}\rangle\Big|_{z^2=0}
 &=& i f_{\pi}P_{\mu}
    \int_{0}^{1} dx\ e^{ix(zP)}\
     \varphi_{\pi}(x,\mu^2)
    \label{eq:PiME}
\end{eqnarray}
describes the transition of a pion $\pi(P)$ 
to a pair of valence quarks $u$ and $d$, 
separated by the (straight) Fock--Schwinger connector ${\cal E}$, 
with corresponding momentum fractions 
$xP$ and $\bar{x}P$, ($\bar{x}\equiv 1-x$).

In order to obtain the pion DA we use a QCD sum rule (SR) approach with
non-local condensates (NLC) \cite{MR86}, 
employing for the scalar and vector condensates the same minimal model
as in~\cite{BM98,BMS01}
\begin{eqnarray}\label{eq:S.V.NLC}
 \langle{\bar{q}(0)q(z)}\rangle
  = \langle{\bar{q}q}\rangle\,
     e^{-|z^2|\lambda_q^2/8}\,;\quad
 \langle{\bar{q}(0)\gamma_\mu q(z)}\rangle 
 = \frac{i\, z_\mu\,z^2}{4}\,
    \frac{2\alpha_s\pi\va{\bar{q}q}^2}{81}\
      e^{-|z^2|\lambda_q^2/8}\,.
\end{eqnarray}
The nonlocality parameter $\lambda_q^2 = \langle{k^2}\rangle$
characterizes the average momentum of quarks in the QCD vacuum and
has been estimated in QCD SRs~\cite{BI82,OPiv88} and on the
lattice~\cite{DDM99,BM02}:
$\lambda_q^2 = 0.45\pm 0.1\text{~GeV}^2$.
For the quark-gluon-antiquark condensates we use
\begin{eqnarray}
\va{\bar{q}(0)\gamma_\mu(-g\widehat{A}_\nu(y))q(x)}&=&
       (y_\mu x_\nu-g_{\mu\nu}(yx))\overline{M}_1(x^2,y^2,(y-x)^2)\nonumber\\
      &+&
      (y_\mu y_\nu-g_{\mu\nu}y^2)\overline{M}_2(x^2,y^2,(y-x)^2)\,,\nonumber
\\
\va{\bar{q}(0)\gamma_5\gamma_\mu(-g\widehat{A}_\nu(y))q(x)}&=&
       i\varepsilon_{\mu\nu yx}\overline{M}_3(x^2,y^2,(y-x)^2)\,,
\vspace{-5mm}\nonumber
\end{eqnarray}
with ($A_{1,2,3}=A_0\left(-\frac32,2,\frac32\right)$)
\begin{eqnarray}
 \overline{M}_i(x^2,y^2,z^2)
  = A_i\int\!\!\!\!\int\limits_{\!0}^{\,\infty}\!\!\!\!\int\!\!
        d\alpha \, d\beta \, d\gamma \,
         f_i(\alpha ,\beta ,\gamma )\,
          e^{\left(\alpha x^2+\beta y^2+\gamma z^2\right)/4}\,.
\end{eqnarray}
The minimal model of nonlocal QCD vacuum suggests 
the following Ansatze
\begin{eqnarray}
 \label{eq:Min.Anz.qGq}
  f_i\left(\alpha,\beta,\gamma\right)
   = \delta\left(\alpha -\Lambda\right)\,
       \delta\left(\beta -\Lambda\right)\,
        \delta\left(\gamma -\Lambda\right)
\end{eqnarray}
with $\Lambda=\frac12\lambda_q^2$ and faces problems 
with QCD equations of motion 
and gauge invariance of 2-point correlator of vector currents.
In order to fulfil QCD equations of motion exactly
and minimize non-transversity of $V-V$ correlator
we suggest the improved model of QCD vacuum with 
\begin{eqnarray}
 \label{eq:Imp.Anz.qGq}
 f^\text{imp}_i\left(\alpha,\beta,\gamma\right)
  = \left(1 + X_{i}\partial_{x} + Y_{i}\partial_{y} + Y_{i}\partial_{z}\right)
         \delta\left(\alpha-x\Lambda\right)
          \delta\left(\beta-y\Lambda\right)
           \delta\left(\gamma-z\Lambda\right)\,,
\end{eqnarray}
where $z=y$, $\Lambda=\frac12\lambda_q^2$ and
\begin{subequations}
\begin{eqnarray}
  X_1 &=& +0.082\,;~X_2 = -1.298\,;~X_3 = +1.775\,;~x=0.788\,;~~~\\
  Y_1 &=& -2.243\,;~Y_2 = -0.239\,;~Y_3 = -3.166\,;~y=0.212\,.~~~
\end{eqnarray}
\end{subequations}
\begin{figure}[t]
 $$\includegraphics[width=0.45\textwidth]{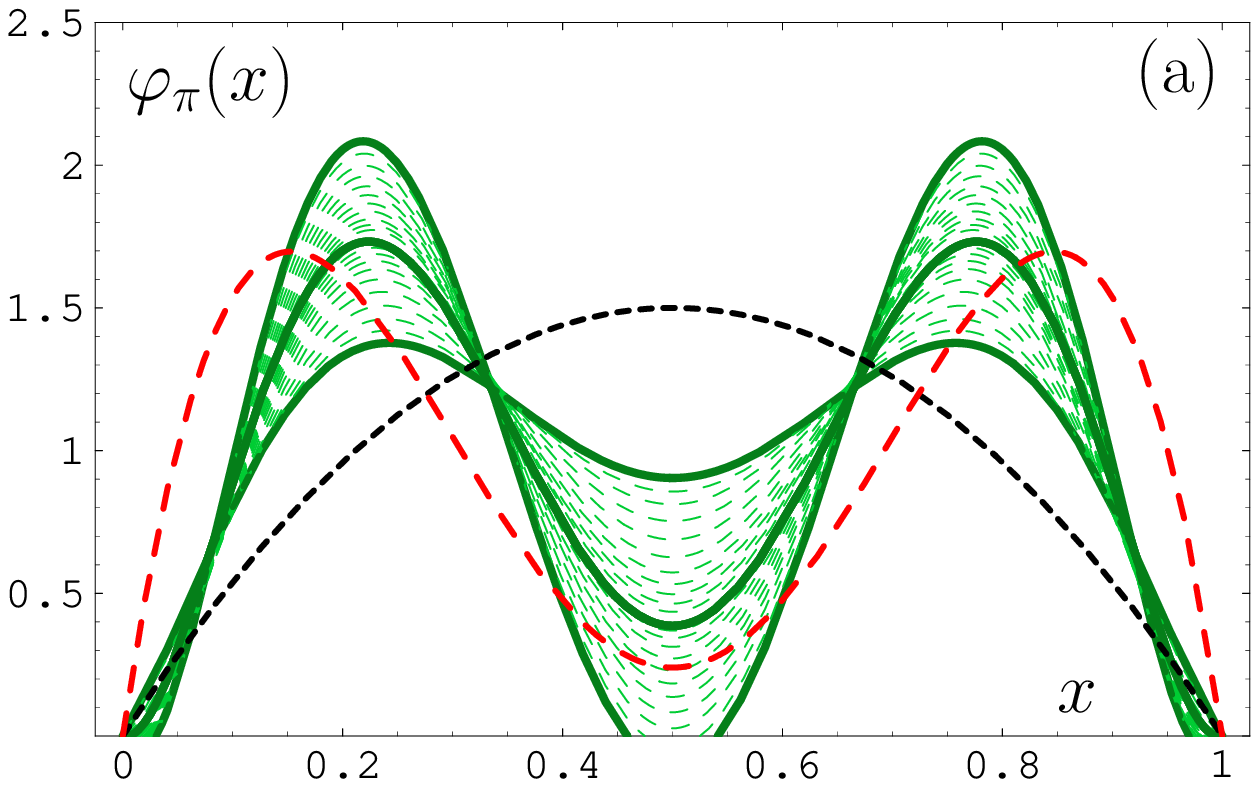}~~~~~~~%
    \includegraphics[width=0.45\textwidth]{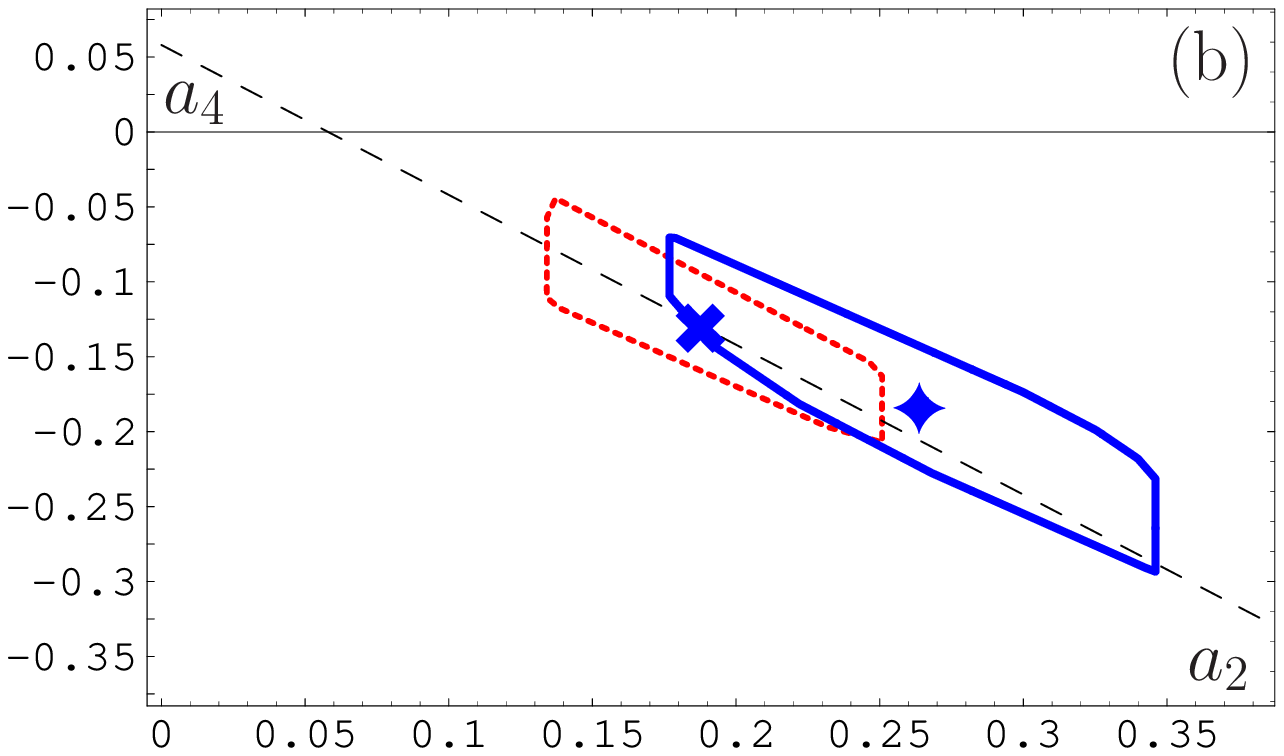}$$\vspace*{-9mm}
   \caption{\label{fig:Pion.DA.Bunch}\footnotesize
    (a) ``Bunch'' of pion DAs extracted from NLC QCD sum rules
    \protect\cite{BP06}. 
    For comparison we show here also the asymptotic DA (dotted line) 
    and Chernyak--Zhitnitsky (CZ) DA~\protect\cite{CZ82} (dashed red line).
    (b) Allowed values of the pion DA parameters $a_2$ and $a_4$ are
    bounded by the solid blue line. 
    Region bounded by the dotted red line represents results obtained 
    in the minimal model~\protect\cite{BMS01}.
    Both panels show results for the value $\lambda_q^2=0.4$ GeV$^2$.}
\end{figure}\vspace*{-1mm}
Then the NLC sum rules for the pion DA produce~\cite{BP06}
a ``\textit{bunch}'' of 2-parameter models at $\mu^2=1.35$ GeV$^2$
(with $\varphi^{\text{as}}(x)\equiv6\,x\,(1-x)$)
\begin{equation}
 \label{eq:2p-Bunch}
  \varphi_\pi(x)
  = \varphi^{\text{as}}(x)
      \left[1 + a_2 C^{3/2}_2(2x-1) + a_4 C^{3/2}_4(2x-1)
      \right]\,,
\end{equation}
shown in Fig.\ \ref{fig:Pion.DA.Bunch}a. 
Allowed values of this bunch parameters $a_2$ and $a_4$
are shown in Fig.\ \ref{fig:Pion.DA.Bunch}b
with coordinates of the central point \BluTn{\ding{70}} to be
$a_2=0.268$ and $a_4=-0.186$. 
These values correspond to $\langle{x^{-1}}\rangle_\pi^{\text{bunch}} = 3.24\pm0.20$,
which is in agreement with the result of an independent sum rule, viz.,
$\langle{x^{-1}}\rangle_\pi^{\text{SR}}=3.40\pm0.34$.

We emphasize here that BMS model~\cite{BMS01},
shown in Fig.\ \ref{fig:Pion.DA.Bunch}b by symbol \BluTn{\ding{54}},
is inside the allowed region dictated by the improved QCD vacuum model.
This means that all the characteristic features of the BMS bunch 
are valid also for the improved bunch:
one can see in Fig.\ \ref{fig:Pion.DA.Bunch}a
that in comparison with CZ model~\cite{CZ82} 
(dashed red line, $a_2=0.56$ and $a_4=0$ at $\mu^2=1$~GeV$^2$)
NLC-dictated models are much more end-point suppressed,
although are double-humped.

\section{NLO light-cone\ sum rules (LCSR), CLEO data and lattice QCD}
The CLEO experimental data~\cite{CLEO98} on $F^{\gamma\gamma^*\pi}(Q^2)$ allow one
to obtain direct constraints on $\varphi_\pi(x)$.
Applying the LCSR approach~\cite{Kho99,SY99}, one can effectively account
for the long-distance effects of a real photon by using quark-hadron
duality in the vector channel and a dispersion relation in $q^2$.

\begin{figure}[b]\vspace*{-3mm}
 \centerline{\includegraphics[width=0.45\textwidth]{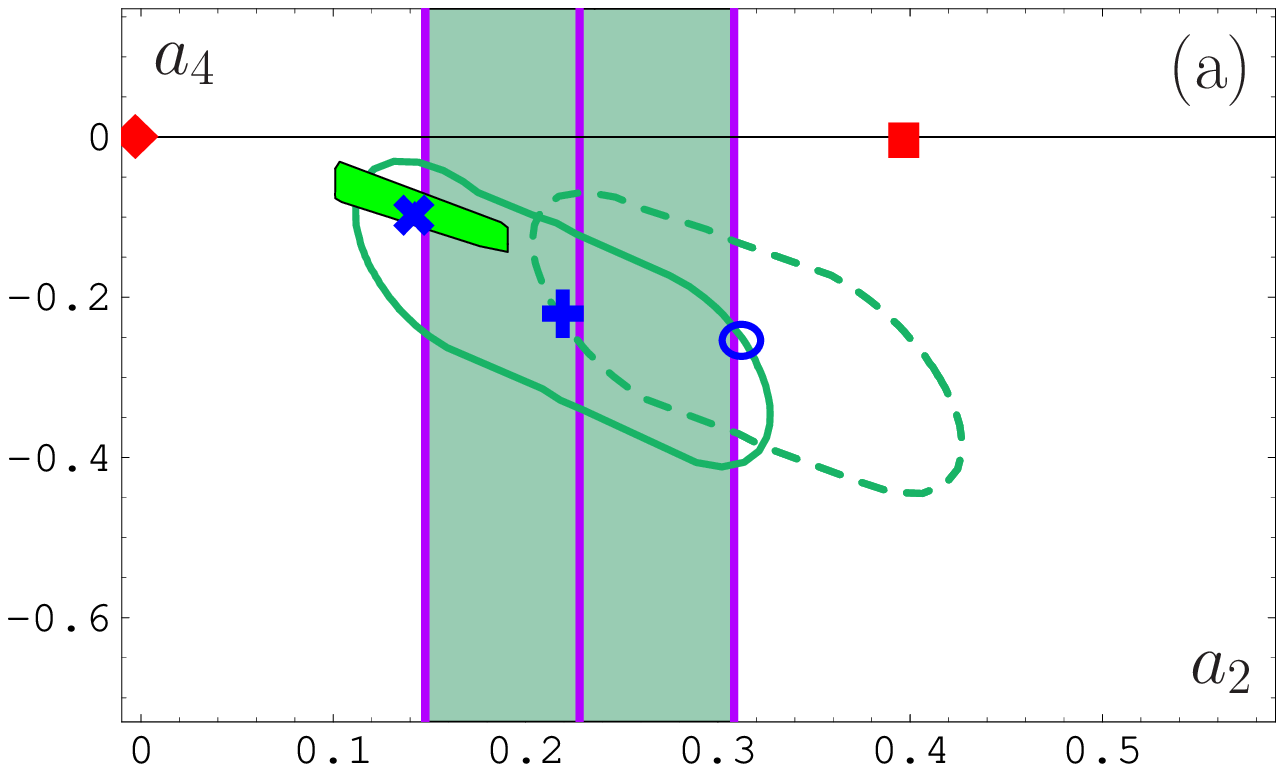}~~~~~~~%
  \includegraphics[width=0.45\textwidth]{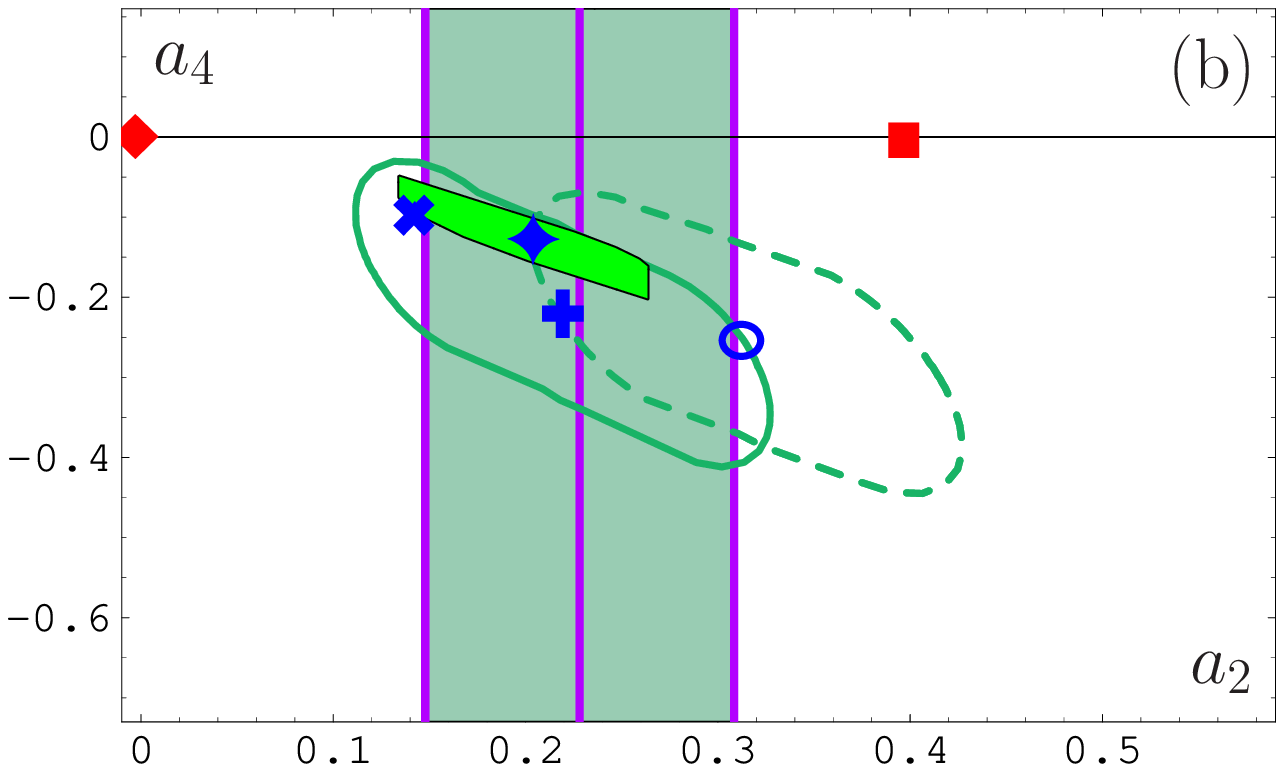}}
   \vspace*{-1mm}
   \caption{Results of the LCSR-based CLEO-data analysis
    on $F_{\pi\gamma^{*}\gamma}(Q^2)$ in comparison with the lattice results
    of \protect\cite{Lat05}, shown as shaded area.
    $1\sigma$-ellipse of~\protect\cite{BMS02} is enclosed by the solid line,
    while the renormalon-based one~\protect\cite{BMS05lat} -- by the dashed line. 
    Panel (a) shows comparison with predictions of the minimal NLC model, 
    whereas panel (b) -- with those of the improved NLC model,
    displayed in both cases as slanted shaded rectangles.
    The displayed models are:
    \RedTn{\ding{117}} -- the asymptotic DA;
    \BluTn{\ding{54}} -- BMS model~\protect\cite{BMS01};
    \BluTn{\ding{70}} -- the central point of our new bunch~\protect\cite{BP06};
    \RedTn{\footnotesize\ding{110}} -- CZ model~\protect\cite{CZ82}.
    All results are evaluated at $\mu^2_\text{SY}=5.76~\text{GeV}^2$
    after NLO ERBL evolution. \label{fig:BMS-lattice}\vspace*{-3mm}}
\end{figure}

In our CLEO data analysis \cite{BMS02}, 
we also used the relation between $\lambda_q^2$ 
and the twist-4 magnitude
$\delta^2_\text{Tw-4} \approx \lambda_q^2/2$ and estimated
$\delta^2_\text{Tw-4}= 0.19 \pm 0.02$ at $\lambda_q^2=0.4$ GeV$^2$.
We used, following the approach of~\cite{Kho99,SY99}, 
the asymptotic model for the twist-4 contribution.
We found that even with a 20\% uncertainty in $\delta_\text{Tw-4}^2$,
the Chernyak--Zhitnitsky (CZ) DA \cite{CZ82} was excluded
\emph{at least} at the \textbf{$4\sigma$}-level, 
whereas the asymptotic DA was off the \textbf{$3\sigma$}-level, 
while the BMS ``bunch'' was inside the $1\sigma$-region~\cite{BMS02},
shown in Fig.\ \ref{fig:BMS-lattice} 
as a solid ovals around the best-fit point (\BluTn{\ding{58}}).

Another possibility, suggested in~\cite{Ag05b}, 
to obtain constraints on the pion DA in the LCSR analysis 
of the CLEO data  -- 
to use for the twist-4 contribution renormalon-based model,
relating it then to parameters $a_2$ and $a_4$ of the pion DA.
Using this method we obtain~\cite{BMS05lat} the renormalon-based constraints 
for the parameters $a_2$ and $a_4$,
shown in Fig.\ \ref{fig:BMS-lattice} in a form of $1\sigma$-ellipses 
(dashed contours) around the corresponding best-fit point (\BluTn{\large$\bm{\circ}$}).

Recently, high-precision lattice measurements of the second moment
$\va{\xi^2}_{\pi} = \int_0^1(2x-1)^2\varphi_\pi(x)\,dx$
of the pion DA were reported 
by two different collaborations~\cite{DelD05,Lat05}.
Both groups extracted from their respective simulations, 
values of $a_2$ at the Schmedding--Yakovlev scale
$\mu^2_\text{SY}$ around $0.24$,
but with different error bars.
It is remarkable that these lattice results are in striking agreement
with the previous~\cite{BMS01} and improved~\cite{BP06} estimates 
of $a_2$ both from NLC QCD SRs 
and also from the CLEO-data analyses---based 
on LCSR---\cite{SY99,BMS02}, 
as illustrated in Fig.\ \ref{fig:BMS-lattice}, 
where the lattice results of~\cite{Lat05}
are shown in the form of a vertical strip, 
containing the central value with associated errors.

Noteworthily, the value of $a_2$ of the displayed lattice measurements
(middle line of the strip) is very close to the CLEO best fit 
in~\cite{BMS02} (\BluTn{\ding{58}}), 
whereas almost all the bunch,
dictated by the improved NLC QCD SRs~\cite{BP06},
is inside the strip.
Moreover, this bunch is completely inside 
the standard CLEO $1\sigma$-ellipse
and partially inside the renormalon-based CLEO $1\sigma$-ellipse.

\section{Conclusions}
So, we can conclude that 
the two-humped and endpoint-suppressed profile of the pion DA 
emerging from the CLEO-data analysis 
is consistent 
with that we have determined independently 
from QCD sum rules with nonlocal condensates~\cite{BMS01,BP06}.
The improvement of the NLC model, 
suggested in~\cite{BP06},
shifts the allowed region and puts it 
just in the intersection 
of the CLEO-data $1\sigma$-regions,
obtained using the asymptotic and the renormalon-based 
models for twist-4 contribution.
Remarkably, this intersection lies almost 
in the center of the recent lattice-QCD strip.

\section*{Acknowledgments}
This investigation was supported in part 
by the Bogoliubov--Infeld Programme, grant 2006,
by the Heisenberg--Landau Programme, grant 2006, 
and the Russian Foundation for Fundamental Research, grant No.\ 06-02-16215.


\end{document}